\documentstyle[12pt,epsfig]{article}
\begin{document}
\title{Resonances and Off--Shell Characteristics
        of Effective Interactions \footnote{Nuclear Physics A in press}}

\author{S.E.  Massen\footnote{Permanent address: Department of Theoretical
Physics, University of Thessaloniki, Thessaloniki 54006, Greece.} $^a$,
S.A. Sofianos $^a$,  S.A. Rakityansky $^a$, S. Oryu $^b$ \\
\\
$^a$ Physics Department, University of South Africa,\\
  P.O.Box 392, Pretoria 0003, South Africa\\
$^b$ Department of Physics, Science University of Tokyo,\\ 
2641 Yamazaki, Noda, Chiba 278-8510, Japan}

\maketitle

\begin{abstract}
The importance of including  experimental resonances in constructing
effective inter--cluster interactions has been investigated. For this,
we  first  address  the question of how to obtain the analytical
properties of the Jost function in regions of physical interest on
the complex  $k$--plane when the potential is given in a tabular
form.
We then employ the  Marchenko inverse scattering method to construct,
numerically, phase equivalent local potentials supporting the
same bound state(s) but having different resonance spectra
which affect the off--shell characteristics of the corresponding
scattering amplitudes. This implies that the inclusion
of the experimental resonances in constructing a potential would
change  its shape, strength, and range which in turn would modify
the bound and scattering wave functions in the interior region.
This is expected to  have important consequences in calculations of
transition amplitudes in nuclear reaction theories, which strongly
depend on the behaviour of the wave functions at short distances. The
influence of  Supersymmetric Transformations on the position and movement of
resonances has also been investigated.\\


\end{abstract}

\section{Introduction}
A central  problem of the theory of nuclear reactions  is to find the
interaction potential between  the colliding nuclei that can describe a wide
range of scattering data and provide the wave functions required. Such a
potential is not simply the result of mapping data to a convenient functional
form. Rather, it has an underlying physical basis in that it is related to
the nucleon--nucleon  interaction as well as to the structure and dynamics of
the interacting nuclei. 

For practical reasons, it is  desirable to express this potential in terms
of quite simple analytic and local form with parameters adjusted to fit a
set of  scattering data. The  phenomenological optical model potentials
\cite{Hod94} are of this class. The most commonly used form is that of a
Woods--Saxon shape  and for light nuclei that of a Gaussian. As
these potentials   fit the scattering cross section well,  the relevant
scattering wave function is asymptotically correct.
Such potentials, however, do  not guarantee that their off--shell
characteristics are sufficiently  good to describe equally well
reactions that depend on the behaviour of the wave function in the
interior region. Such reactions are, for example, the electro--disintegration
and photo--disintegration processes which depend crucially  on the
wave function at all distances.

When a nucleus--nucleus interaction is constructed theoretically directly
from a nucleon--nucleon potential, it is usually nonlocal and quite
complicated. Such potentials can be obtain, for example, by using the
Resonating Group Method (RGM) in which the inter--cluster interaction
is constructed via the Pauli antisymmetrization \cite{RGM}. To study
the resulting nonlocalities one resorts to a construction of
an equivalent local potential (ELP) using various methods. In this
way certain off--shell characteristics of the interaction can be revealed.
For instance, the use of Fiedeldey's Wronskian
method \cite{F-ELP,FS-ELP} to construct an energy dependent ELP for the
nucleon--$\alpha$ RGM nonlocal interaction revealed a repulsion in the
interaction region \cite{S-ELP} which suggests the possible
existence of resonances.

Another way of constructing local nucleus--nucleus potentials is the use of
inverse scattering techniques. The potential in this case is directly
related to the available information of the scattering phase shifts and
of bound states.  In the inverse scattering method at a specific partial
wave $\ell$ (fixed--$\ell$ inversion)   \cite{Marchenko,Chadan,Newton}
and in the absence of
bound states, the constructed potential is unique. However, when bound states
are present one may construct an infinite number of potentials
which are spectrum and phase equivalent at all energies.
This is achieved by choosing arbitrarily the bound states normalization
constants which are not available from experiments.  However, as
we shall show in this paper, the asymptotic normalization constants
determine the distribution of resonances in the $k$--plane and vice versa.
This emphasizes the importance of taking resonances into account in
constructing effective interactions which are usually ignored.

One of the main reasons for omitting the resonances is  the lack of
experimental information on their positions and widths especially for broad
resonances. In the past this was aggravated  by  the absence of an exact and
yet simple method to study the analytical properties of the corresponding
amplitude which could facilitate their incorporation into  the potential
construction. Recently, however, a new method for direct calculation of the
Jost function in the complex $k$--plane, has been  developed
\cite{res1,res2,res3,res4}. Within this method,  the bound, resonant, and
scattering states can be found by locating  the Jost function zeros in  the
appropriate domain of the $k$--plane. It is based on a combination of the
complex coordinate rotation with the variable--constant method used to
replace the Schr\"odinger equation by an equivalent system of linear
first--order differential equations. Since this method is both exact and
quite simple, it could be used, together with  the fitting of phase shifts,
to construct a realistic potential which reproduces  resonance 
poles  as well.

Complex rotation of the coordinate requires knowledge of the potential
off the real axis in the $r$--plane. This poses no problem when the
potential is given in an analytic form. However, there are cases, such as the
aforementioned inverse scattering method, in which  the resulting potential
is given as a tabulated set of values versus radial points. In this paper,
we show how the Jost function method of Refs. \cite{res1,res2,res3,res4} can
be extended to deal with such potentials,  and then apply it to
study the resonances and their effects on the off--shell characteristics
of phase equivalent potentials  obtained numerically by  the Marchenko
inversion and in Supersymmetric (SUSY)  transformations.

In Sec. II we will describe our formalism by briefly recalling
the Marchenko inversion method, the relevant SUSY transformations,
and the  exact method of obtaining resonances and Regge trajectories.
Our results are presented in Sec. III and our conclusions are given in
Sec. IV. In the Appendix we present some technical details related
to the inverse method and the Jost function.
\section{Formalism}
\subsection{Inverse scattering method}
\label{IIA}
In the Marchenko inversion scheme \cite{Marchenko,Chadan,Newton}
the potential $V_\ell(r)$, for each partial wave $\ell$,  is obtained from
the relation
\begin{equation}
	V_\ell(r)=-2\frac{{\rm d} K_\ell(r,r)}{{\rm d}r}
\label{Vr}
\end{equation}
where  the function $K_\ell(r,r')$ satisfies the Marchenko fundamental
integral equation

\begin{equation}
	K_\ell(r,r')+{\cal F}_\ell(r,r')+\int_r^\infty
		K_\ell(r,s){\cal F}_\ell(s,r'){\rm d} s=0\ .
\label{March}
\end{equation}


%
The kernel  ${\cal F}_\ell(r,r')$ of this equation is related to the
$S$--matrix $S_\ell(k)$, and thus to experiment,  via
\begin{equation}
     {\cal F}_\ell(r,r')=\frac{1}{2\pi} \int_{-\infty}^\infty
     h_\ell^{(+)}(kr)\left[1-S_\ell(k)\right]h^{(+)}_\ell(kr'){\rm d}k-
      \sum_{n=0}^{N_b-1}
     A_{n \ell}h^{(+)}_\ell(b_n r)h^{(+)}_\ell(b_n r')\ ,
\label{frr}
\end{equation}
where $h^{(+)}_\ell(z)$ is the Riccati--Hankel function, $N_b$ is the number
of bound states,  and $A_{n \ell}$  is the asymptotic normalization constant
\cite{Chadan,Newton} for the n${th}$ bound state with  energy
$E_b^{(n)}=-\hbar^2b_n^2/2\mu $ where $ib_n$ lies on the positive
imaginary axis of the $k$--plane. The  $A_{n \ell}$  can be expressed
in terms of the relevant Jost solution ${\rm f}_\ell^{(+)}(k,r)$
\begin{equation}
	A_{n \ell}^{-1}=\int_0^{\infty}
	      \left[ {\rm f}^{(+)}_\ell(ib_n,r)\right]^2{\rm d}r\,.
\label{asym1}
\end{equation}
The $S$--matrix needed in Eq. (\ref{frr}) can be parametrized using the
convenient rational ansatz,
\begin{equation}
	  S_\ell(k)=\prod_{i=1}^{\infty} \frac{k+ a_i}{k- a_i}\ ,
\label{Smat2}
\end{equation}
{\em i.e.} using an infinite number of simple poles and zeros. In practice
the number of $a_i$ in the above formula is limited to $N$ which is
sufficiently large to reproduce the input scattering phase shifts and bound
states. The conventions  and a method to evaluate these $N$ constants are
described in  Appendix A. The  rational   parametrization of the $S$--matrix
reduces the Marchenko inverse scattering procedure to an algebraic
problem since the kernel of Eq. (\ref{March}) becomes separable;
this can easily be seen if one performs the integration in
Eq. (\ref{frr}) using the residue method.

In the absence of bound states  the above scheme is unique, {\em i.e.}, once
a good fit to the data from $(0,\infty)$ is achieved, one and only one
potential can be obtained. In the presence of bound states, however, the
potential is not unique as it depends on the
choice of the asymptotic normalization constants  $A_{n\ell}$ which are not
provided by experiment. When  $A_{n\ell}$ are chosen according to
Eq. (\ref{asym1}) or, equivalently, obtained via the Jost function
$f_\ell(k)$,
\begin{equation}
	     A_{n\ell}=\left[i \frac{f_\ell(-k)}
	{{\rm d}f_\ell(k) / {\rm d}k}\right]_{k=ib_n}\ ,
\label{asym2}
\end{equation}
the resulting potential is unique and  has the  shorter range \cite{Newton}.
Any other choice can lead to an equivalent local potential which reproduces
the same on--shell data but it has  different shape and range.

From Eq. (\ref{asym2}) it is clear that values of the
normalization  constants are determined by the explicit form of the Jost
function, if known.  The $A_{n\ell}$
can be easily obtained using Eq. (\ref{asym2}) where the Jost function
is parametrized using also a rational representation
\begin{equation}
	f_\ell(k)=\prod_{i=1}^\infty \frac{k-\alpha_i}{k-\beta_i}\ .
\label{F1}
\end{equation}
A way to select the alphas and betas, for a finite number of terms used
in the parametrization  (\ref{Smat2}) or (\ref{F1}),
is described in the Appendix.

The Jost function thus constructed approximates well the exact
Jost function on a segment of the real axis of the $k$--plane. From this
one should expect that it would be a good approximation off the real
axis as well. Indeed, if two analytical functions
coincide on any arc of a continuous curve, they coincide everywhere in the
region of their analyticity \cite{Morse}.

There are at least two problems concerning such a parametrization. Firstly,
the function (\ref{F1}) actually does not coincide with the exact Jost
function on the real axis but can only be a good numerical
approximation at certain points. How fast the deviations are growing when
one  moves away from the real axis is  not known. Secondly, since instead
of an infinite number of terms in Eq. (\ref{F1}), we have to truncate the
product to a finite number of terms, not all $\alpha_i$ correspond to zeros of
the exact Jost function. In other words, the fitting procedure employed to
evaluate these parameters, may generates a number of unphysical
zeros and  poles which could be simply  an artifact of the truncation.
%
\subsection{Supersymmetric Transformation}
\label{IIB}
The inter--cluster interaction in nuclear physics is usually characterized
by a repulsion at short distances. Alternatively, the potential can be
made deep enough to sustain  deep bound states
which simulate the role of the so-called  Pauli Forbidden States (PFS)
appearing in  the RGM theory which are not physical.  This means
that the Levinson's theorem,
        $\delta_0(0)-\delta_0(\infty)=\pi$,
is fulfilled for this system. These states, however, pose a problem when
the potential is intended for used in few--body calculations as they
generate a scattering amplitude with different off--shell characteristics.
Thus one resorts  to the removal  of the PFS using a subtraction
technique \cite{Kuku} that results in a shallow  potential.
A rigorous  relation between  deep and shallow potentials, however, has been
established by Baye \cite{Bay87} by  using two successive SUSY
transformations, the first in order to remove
the ground state and the second to restore phase equivalence.
Alternatively the ground state can be removed  using the Marchenko
inverse scattering method and assigning a zero value
to the corresponding asymptotic normalization constant. The resulting
potential is also unique \cite{Sof90}  coinciding with the one obtained
from SUSY transformations. In what follows, the main equations of the
supersymmetric transformation will be briefly recalled (details
can be  found, for example, in Refs. \cite{Bay87,Suk85}).

The supersymmetric transformation of a given potential $V^{(0)}_\ell$ that 
has an undesirable ground state at $E=E_b^{(0)}=-\hbar^2 b_0^2/2\mu$, 
results in a
phase--equivalent potential $V^{(2)}_\ell$ which does not sustain this state.
Two consecutive transformations,
\begin{equation}
\label{twosteps}
    V^{(0)}_\ell\Longrightarrow V^{(1)}_\ell\Longrightarrow V^{(2)}_\ell\ ,
\end{equation}
are required for this purpose. The first one eliminates the ground state of
the original potential, but also changes the phase shifts. Phase equivalence
is then restored by a second SUSY transformation.
The first step of the transformation (\ref{twosteps}),
\begin{equation}
\label{firststep}
       V^{(1)}_\ell(r)=V^{(0)}_\ell(r)-2\frac{{\rm d}^2}{{\rm d}r^2}
       \ln\Psi_{0\ell}^{(0)}(ib_0,r)\ ,
\end{equation}
requires the normalized wave function $\Psi_{0\ell}^{(0)}(ib_0,r)$ of the
ground state to be removed. The second step is implemented via
\begin{equation}
\label{secondstep}
       V^{(2)}_\ell(r)=V^{(0)}_\ell(r)-2\frac{{\rm d}^2}{{\rm d}r^2}
       \ln\left[\Psi_{0\ell}^{(0)}(ib_0,r)\Psi_{\ell}^{(1)}(ib_0,r)
       \right]\ ,
\end{equation}
where the wave function $\Psi_{\ell}^{(1)}(ib_0,r)$ is the
solution of the Schr\"odinger equation with the potential
$V^{(1)}_\ell(r)$ at the same energy $E=E_b^{(0)}$ and has the
asymptotic behaviour
$$
       \Psi_{\ell}^{(1)}(ib_0,r)\ \mathop{\longrightarrow}_{r\to\infty}\
       \exp(b_0r)\ .
$$
The well--known  $1/r^2$   behaviour of the transformed potential
$V^{(2)}_\ell(r)$ at short distances can  easily be deduced from
Eq. (\ref{secondstep}). Indeed, both wave functions
$\Psi_{0\ell}^{(0)}(ib_0,r)$ and $\Psi_{\ell}^{(1)}(ib_0,r)$
near the origin have the behaviour $\sim r^m$ with some integer $m$,
which makes the second term of this equation proportional to
$ \sim 1/r^2$. 

Finally, we note that the SUSY potentials are intrinsically related  to
inverse scattering theory \cite{Suk85}  which can be used to
generate energy--independent shallow potentials from energy--dependent deep
potentials \cite{Fied90}. For this, one has to calculate first the phase
shifts at all energies and then employ the Marchenko scheme to construct
an $\ell$--dependent potential from which a  unique shallow
potential can be obtained  either by super transforming twice or, as
mentioned earlier, by assuming that the corresponding asymptotic
normalization constant is zero.
\subsection{Exact method for locating resonances}
\label{IIC}
The method we employ here for locating potential resonances belongs to
the class of so-called complex energy methods which are based on a rigorous
definition of resonances, namely, as  zeros of the Jost function.
Unlike most of the other methods of this class, which involve an expansion
of the resonant wave function in terms of square--integrable functions, our
method is based on a direct calculation of the Jost function at complex $k$
by integrating exact differential equations equivalent to the Schr\"odinger
equation.

By locating a complex zero $k_r$ of the Jost function in the fourth
quadrant of the momentum plane, we can obtain, at the same time, the
resonant energy $E_r$ and width $\Gamma$  from the simple relation
$$
       E_r-\frac{i}{2}\Gamma=\frac{\hbar^2}{2\mu}k_r^2\ .
$$
It is clear that the accuracy of $E_r$ and $\Gamma$ is related
to the accuracy in  calculating the Jost function itself.
The latter  is obtained, in our method, from the asymptotic value
of a  function $F_\ell^{(-)}(k,r)$ ,
\begin{equation}
\label{flimF}
       f_\ell(k)=\lim_{|r|\to\infty}F_\ell^{(-)}(k,r)\ ,
\end{equation}
which, together with its partner
$F_\ell^{(+)}(k,r)$, obeys the system of first order differential equations
\begin{eqnarray}
\nonumber
\partial_rF_\ell^{(+)}(k,r)&=&\phantom{+}\displaystyle{
	  \frac{h_\ell^{(-)}(kr)}{2ik}
	  V_\ell(r)\left[
	  h_{\ell}^{(+)}(kr)F_\ell^{(+)}(k,r)+
	  h_{\ell}^{(-)}(kr)F_\ell^{(-)}(k,r)\right]}\ ,\\
\label{fpmeq}
&&\\
\nonumber
\partial_rF_\ell^{(-)}(k,r)&=&-\displaystyle{
	  \frac{h_\ell^{(+)}(kr)}{2ik}
	  V_\ell(r)\left[
	  h_{\ell}^{(+)}(kr)F_\ell^{(+)}(k,r)+
	  h_{\ell}^{(-)}(kr)F_\ell^{(-)}(k,r)\right]}
\end{eqnarray}
the boundary conditions being
\begin{equation}
\label{bcondF}
	  F_\ell^{(\pm)}(k,0)=1\ .
\end{equation}
The origin of the relation (\ref{flimF}) becomes clear if one notices that
the sum of the products of the auxiliary functions $F_\ell^{(\pm)}(k,r)$
with the Riccati--Hankel functions,
\begin{equation}
\label{ansatz}
	  \phi_\ell(k,r)=\frac12\left[
	  h_{\ell}^{(+)}(kr)F_\ell^{(+)}(k,r)+
	  h_{\ell}^{(-)}(kr)F_\ell^{(-)}(k,r)\right]\ ,
\end{equation}
obeys the Schr\"odinger equation. The function $\phi_\ell(k,r)$ is
the so-called {\it regular solution} which vanishes near $r=0$ exactly
like the Riccati--Bessel function, {\it i.e.}
\begin{equation}
\label{regconF}
	      \lim_{r\to 0} \phi_\ell(k,r)/j_\ell(kr) = 1\ .
\end{equation}
The Jost function, Eq. (\ref{flimF}), can be obtained by comparing
the asymptotic behaviour of the regular solution,
\begin{equation}
\label{assansatz}
	  \phi_\ell(k,r)=\mathop{\longrightarrow}_{r\to\infty}\frac12\left[
	  h_{\ell}^{(+)}(kr)f_\ell^{*}(k^*)+
	  h_{\ell}^{(-)}(kr)f_\ell(k)\right]\ ,
\end{equation}
with Eq. (\ref{ansatz}) expressed  in terms of the auxiliary functions
$F_\ell^{(\pm)}(k,r)$.

In Ref. \cite{res2} it has been shown that  the limit (\ref{flimF})
exists for all complex $k$ for which
\begin{equation}
\label{limcond}
	  {\rm Im\,}kr\ge 0\ .
\end{equation}
If $r$ is real, the condition (\ref{limcond}) is only satisfied for bound
and scattering states but not for resonances. Therefore, to calculate
$f_\ell(k)$ in the fourth quadrant we make the complex rotation of
the coordinate in Eqs. (\ref{fpmeq}),
\begin{equation}
\label{rot}
	r=x\exp(i\theta)\, ,\qquad x\ge 0\, ,
		\qquad 0\le\theta<\frac{\pi}{2}\ ,
\end{equation}
with a sufficiently large $\theta$.

The above scheme works extremely well in locating bound, scattering, and
resonant states as well in finding  Regge poles and trajectories when
the potential $V_\ell(r)$  is central, short range and it is given in
analytic form. A generalization to non--central, multi--channel, and
Coulomb--tailed  as well as to singular potentials can be found in Refs.
\cite{res1,res2,res3,res4}. However, in this work we are concerned
with potentials given in a tabular form and the  question of how to
handle such potentials will be discussed  when presenting our results.
\section{Results}
In the present work the potentials were generated  either by inversion
or by SUSY transformations and hence in both cases they are available in
a tabular form. To make an analytic continuation of them into the
first quadrant of the complex $r$--plane, needed for the complex
rotation, we fitted the potentials on the real axis by simple
analytical forms with adjustable parameters and then considered $r$ in
these forms as a complex variable. Such an approach to analytic continuation
is based on a theorem of the complex analysis which states that if
a function is analytic in a region and vanishes along
any arc of a continuous curve in this region, then it must vanish
identically in this region \cite{Morse}. The obvious corollary of this
theorem is that if two functions coincide on a curve, they coincide everywhere
in the region of analyticity. Therefore, the analytical form which
coincides with the tabulated function on the real axis should reproduce
this function off the real axis as well.  The question then arises what
if the potentials coincide within numerics.
In other words, we want to know whether  small numerical
deviations in the potential on the real axis  generate perceptible
deviations of the position of the resonances. We have
investigated this situation first by assuming the following
analytic potential
\begin{equation}
	V(r)=5 \exp\left[-0.25(r-3.5)^2\right]-8\exp(-0.2r^2)
\label{som0}
\end{equation}
where the strength parameters are given in MeV and $r$ in fm. The
reduced mass $\mu$ is  such that $\hbar^2/2\mu=1/2$\,MeV\,fm$^2$.
The resonances and Regge trajectories of this potential were
investigated in Ref. \cite{res2}. We then fitted the $N$ points
$V(r_i)$, $i=1,2,\cdots, N$ using the ansatz
\begin{equation}
	V_{\rm fit}(r)=\sum_{n=1}^{N_1}a_n \exp\left[-b_n(r-c_n)^2\right]
                   +  \sum_{n=1}^{N_2}d_n \exp(-e_n r^2)
\label{sum1}
\end{equation}
where the  parameters were obtained using  the MERLIN minimization program
\cite{MERLIN}. The minimization was stopped when the least square error on
120 points and $N_1=5$ and $N_2=4$ was of the order of 10$^{-4}$ -- a
typical accuracy in these cases. This means that the fit to the analytic
potential was between the third and fourth decimal in the whole region.
With such a fit, all resonances found in Ref. \cite{res2} were
recovered  within three to five decimal points.  For comparison,
the energies and widths for few of
them (the lowest resonances in each partial wave) are given in Table
\ref{recov}.  Obviously the accuracy can be improved as the fit to the
potential improves.  One further comment is necessary: The form factors
used for the fit  restrict the use of a rotation to only a certain region.
In this respect  the use of splines is not suitable at all as they
diverge for all angles of rotation. 

These test calculations show that the Jost function method based on the
complex rotation of the coordinate, is applicable and retains its
effectiveness even when the potential is given numerically on the real axis.
We can therefore use it to study the importance of incorporating
physical two--cluster  resonances in constructing a potential. This can
be easily investigated by constructing ELP's for a specific partial
wave via inversion as described above and studying the implications of the
implanted resonances. For this we use the nucleon--$\alpha$
potential of Dubovichenko and Dzhazairov--Kakhramanov \cite{Dubo}
for the $\ell=0$ partial wave
\begin{equation}
	V(r)=-V_0\exp({-\alpha r^2})
\label{dubo}
\end{equation}
where $V_0= 55.774$\,MeV and $\alpha=0.292$\,fm$^{-2}$. This is a deep
potential that sustains an  unphysical PFS state at $-9.73058$\,MeV. This
means  that the Levinson's theorem,
        $\delta_0(0)-\delta_0(\infty)=\pi$,
is fulfilled for this system. At large distances the radial wave function
decays exponentially,
$$
      u_0(r)\mathop{\longrightarrow}_{r\to\infty}A_s\exp(-b_0r)\ ,
$$
and  the asymptotic normalization constant was found to be
$A_s=6.1603 \,{\rm fm}^{-1/2}$.

\begin{figure}[t]
\vspace*{-1cm}
\centerline{\epsfig{file=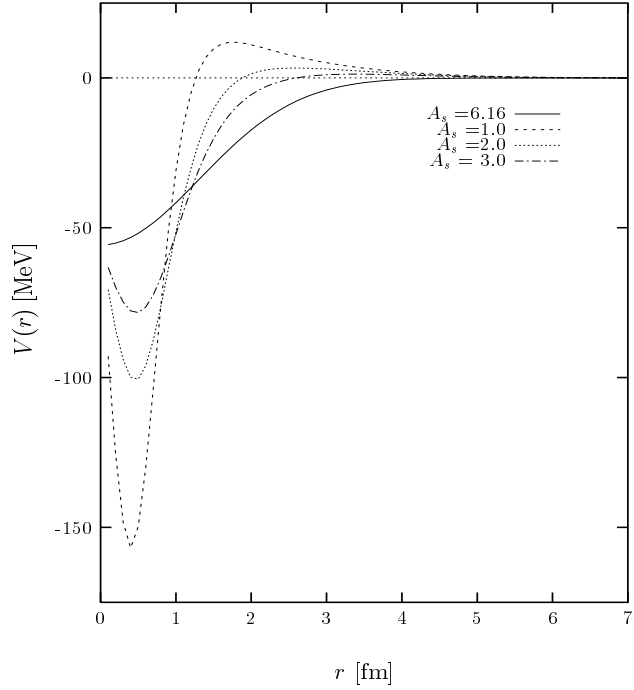,height=10cm,width=10cm}}
\caption{
Bound state and phase equivalant potentials for the 
nucleon--$\alpha$ interaction. These potentials generate different 
resonance spectra.}
\label{Vs}
\end{figure}

By varying the asymptotic normalization constant we obtained a set of
potentials which were fully phase shift and bound state equivalent but have
different number of potential resonances. These potentials are shown in
Fig.~\ref{Vs}. It is seen that for values of  $A_s$ less than the choice
given by (\ref{asym1}) ($A_s=6.1603$),  a hump appears in the interaction
region which is higher as  $A_s$ becomes smaller while at the same time the
well becomes deeper. For values larger than 6.1603 the potential is also of
long range but without a hump. In the extreme case of $A_s=0$ the potential
becomes repulsive at all distances. This means that as $A_s\to 0$ resonances
are generated and  their appearance and position depend on the specific
choice of $A_s$.

\begin{figure}[b]
\vspace*{-1cm}
\centerline{\epsfig{file=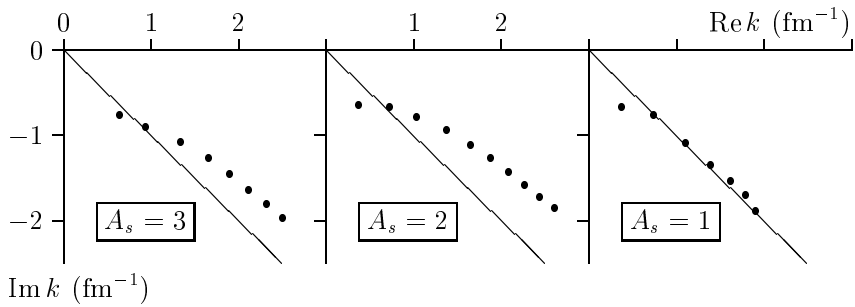,height=6cm,width=15cm}}
\vspace*{-1cm}
\caption{Distribution of the Jost function zeros (filled circles)
 in the complex $k$--plane for the $S$--wave N--$\alpha$ potential for
three different values of the asymptotic normalization constant $A_s$.
The diagonal of the fourth quadrant represents the threshold boundary
${\rm Re\,}E=0$.
}
\label{a3a2a1}
\end{figure}

Using the ansatz (\ref{sum1}) we fitted these  potentials,
with $N_1=5$ and $N_2=3$, via the MERLIN code \cite{MERLIN} 
the accuracy being again within
a fourth decimal at all points meaning that corresponding
accuracy in  reproducing the phase shifts was better than 0.0001 of a 
degree.  We employed  the analytical representations of these
phase--equivalent potentials to locate the zeros of the Jost function 
in the fourth quadrant of the $k$--plane.
The original potential (\ref{dubo}), which is also a member of our set of
the phase--equivalent potentials, does not generate any physical resonances.
All the zeros of the Jost function, which we found for this potential, are
situated below the diagonal of the fourth quadrant of the $k$--plane and,
therefore, represent sub--threshold resonances which are unphysical. The
growth of the potential barrier when $A_s$ decreases, indicates that some
physical resonances should appear. In other words, when $A_s$ becomes
smaller some of the Jost function zeros should move up to the area above the
threshold line. When, however, $A_s$ is too small, the barrier
transforms into a strong  repulsive core, and the resonances should
disappear. This can be seen in Fig. \ref{a3a2a1} where we present the
distributions of the Jost function zeros for three phase--equivalent
potentials corresponding to $A_s$ equal to 3, 2 and 1. It is clear that the
choice of $A_s$ determines positions of the resonances and vice versa.

\begin{figure}[t]
\vspace*{-2cm}
\centerline{\epsfig{file=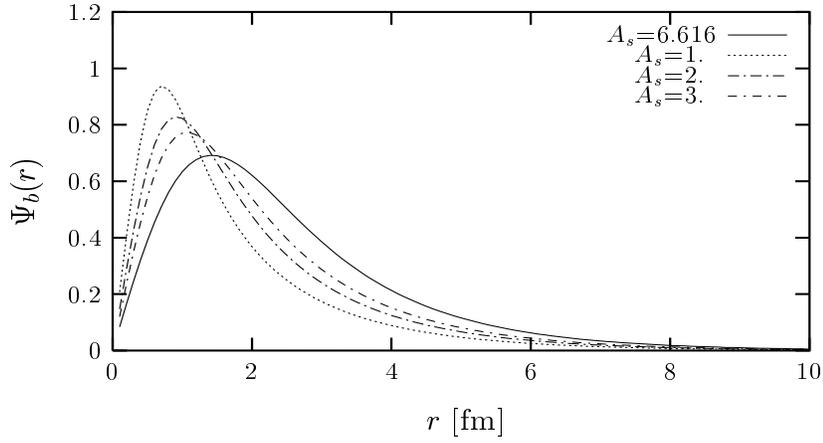,height=13cm,width=13cm}}
\vspace*{-5cm}
\caption{Bound state wave functions generated by the potentials of Fig. 1}
\label{bs-wf}
\end{figure}

\begin{figure}[b]
\centerline{\epsfig{file=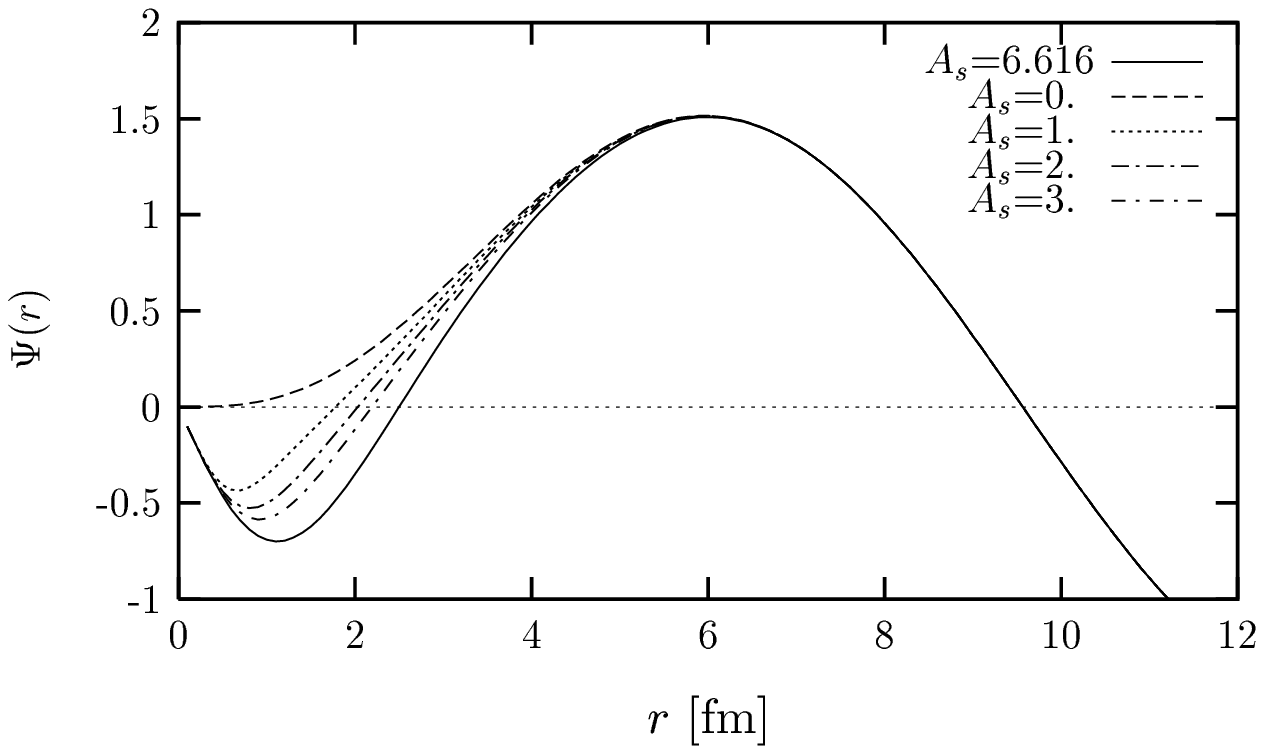,height=13cm,width=13cm}}
\vspace*{-5cm}
\caption{Scattering wave functions generated by the potentials of Fig. 1
  for the center of mass energy $E=5$\,MeV. The $A_s=0$  corresponds 
to the shallow $V^{(2)}$ potential that generates a nodeless 
wave function at short distances. 
}
\label{scat-wf}
\end{figure}

At a first sight one can argue that the zeros of the Jost function
practically have no effect on the scattering processes because they are far
away from the real axis. These potentials, however,  generate bound and
scattering wave functions which have a different behaviour  in the interior
region. The  bound state wave functions are shown in Fig. \ref{bs-wf}
while the scattering wave functions,  for center of mass energy $E=5$\,MeV,
are plotted in Fig. \ref{scat-wf}. The nodeless wave function for the SUSY
potential $V^{(2)}(r)$ is also shown in the latter. Since the interior region
(within few fm) is of  importance in describing various nuclear reactions,
the existence and distribution of resonances cannot be ignored when 
the reaction observables are calculated. These differences are also a 
source of off shell differences in
the corresponding scattering matrices which are manifested in three-- and
four--cluster calculations.

As another example, we consider  the  $\alpha$--$\alpha$ local potential 
of Buck {\it et al.} \cite{Buck77}
\begin{equation}
\label{buckpot}
      V(r)=-V_0\exp(-\alpha r^2)+\frac{4e^2}{r}{\rm erf}(\beta r)\ ,
\end{equation}
with $V_0=122.6225$\,MeV, $\alpha=0.22$\,fm$^{-2}$, and
$\beta=0.75$\,fm$^{-1}$. This  potential sustains, in the $\ell=0$ partial
wave, two unphysical deep bound states at $-72.78$\,MeV and $-22.28$\,MeV and
a resonance at 179.22\,keV with $\Gamma=0.94905$\,keV. The position of the
resonance is quite different from the values of 92.12$\pm$0.05\,keV and
$\Gamma=5.80$\,eV given by Buck {\em et al.} \cite{Buck77}). A possible
reason for such a discrepancy is that in locating these spectral points we
included the Coulomb tail of the second term of Eq.~(\ref{buckpot}) in an
exact way as was proposed in Refs. \cite{res1,res2}.
We are interested about the movement of this resonance when we eliminate 
one of the unphysical bound states using the SUSY transformation. To this end, the
numerically obtained SUSY potential $V^{(2)}$ was fitted using the ansatz
\begin{equation}
	V_{\rm fit}(r)=\sum_{n=1}^{5}a_n \exp(-b_n r^2)/r^2
		    + \sum_{n=1}^{5}c_n \exp(-c_n r^2)
\label{sum2}
\end{equation}
with the accuracy of the fit of 450 points being better than 10$^{-5}$. Using
this analytic potential, we found a zero of the $S$--wave Jost function at
$k=0.10383 - i0.77289\times10^{-5}$\,fm$^{-1}$ ({\em i.e.} at $E=0.11261
-\displaystyle\frac{i}{2}0.33532\times10^{-4}$\,MeV) which is significantly
different from the position of the $S$--wave resonance given in
Table~\ref{bucktable}. In view of the above discussion this is not surprising
since the SUSY transformation drastically changes the asymptotic
normalization constant for the ground state, to zero.
\section{Conclusion}
We demonstrated that for a potential given numerically the  analytic
properties of the corresponding Jost function in the complex $k$--plane can
be obtained via fitting the potential by any analytic form that allows a
complex rotation into the first quadrant of the complex $r$--plane. The
scattering observables, the bound states, and the potential resonances can be
calculated with a sufficient accuracy which is improved with an improved fit
to the potential.

The phase shifts and therefore the on-shell $S$--matrix which are extracted
from experimental data on the real $k$--axis, contain information about
resonances in an indirect way. The phase-shifts ``feel'' the existence of
resonances only when they are close to the real axis (narrow resonances). The
broad resonances, however, remain ``unnoticed'' by the phase-shifts and
therefore a potential which is based on them, generates an $S$--matrix
without the corresponding poles. However, even extremely broad resonances
affect the behaviour of the physical wave function at short distances. This
implies that an information on the distribution of resonances can be a clue
for making a correct choice among very different potentials  generating the
same phase-shifts and the same on-shell $S$--matrix.

Such information can, in principle, be obtained from various inelastic
processes. For example, photo-excitation of a nucleus and its subsequent
decay in two fragments $A$ and $B$ can reveal $AB$--resonances which are not
``visible'' in elastic $AB$--scattering. When an effective potential is
constructed using not experimental but theoretical phase--shifts, additional
effort to locate broad resonances would exclude ambiguities (non-uniqueness)
of the effective potential. This is the case, for example, with the RGM
theory which can produce  very complicated (nonlocal) potentials.
To apply such a potential in realistic calculations, one usually calculates
RGM phase-shifts and, using them, constructs a simple effective
(phase-equivalent) potential.

In summary, when constructing a realistic effective interaction, it is
important to take into account physical resonances. In the case of an
energy-- and $\ell$--independent potential, the resonances at all partial
wave must be incorporated. This will guarantee that a reliable  local
potential is obtained that generates a transition matrix which has correct
off--shell behaviour.


\section*{acknowledgements}
We gratefully acknowledge financial support from the Foundation 
for Research Development of South Africa and the Science 
University of Tokyo. Useful discussions with P.E. Hodgson 
are also acknowledged.

\newpage
\appendix
\section{ Distribution of the $S$--function poles in the complex $k$--plane}
The $S$--function is defined in terms of the Jost functions  by
\begin{equation}
	S_\ell(k)=\frac{f^*_\ell(k^*)}{f_\ell(k)}
\label{Sk0}
\end{equation}
The $f_\ell(k)$ can be parametrized using the  rational form
\begin{equation}
        f_\ell(k)=\prod_{i=1}^N \frac{k-\alpha_i}{k-\beta_i}
\label{Fk0}
\end{equation}
implying that it has $N$ simple poles at $k=\beta_i$ and N simple zeros
at $k=\alpha_i$. The $f(k)$ is  a well defined function in the upper-half
$k$--plane. The poles $\beta_i$ are therefore situated in the lower 
$k$--plane.
In contrast, the $\alpha_i$ can be in both planes. Those in the upper half
$k$--plane are denoted  by $\tau_i$ and those in the lower by $\sigma_i$.
For the $S_\ell$--function we have
\begin{equation}
          S_\ell(k)=\prod_{i=1}^N \frac{k+\alpha_i}{k-\alpha_i}
		\frac{k-\beta_i}{k+\beta_i}
\label{Sk1}
\end{equation}
or
\begin{equation}
          S_\ell(k)=\prod_{m=1}^{N_\sigma}\frac{k+\sigma_m}{k-\sigma_m}
                \prod_{n=1}^{N_\tau}\frac{k+\tau_n}{k-\tau_n}\prod_{i=1}^N
                \frac{k-\beta_i}{k+\beta_i}\, , \quad N_\sigma +N_\tau =N
\label{Sk2}
\end{equation}
Setting $\xi_i=-\beta_i$ (Im\,$\xi_i>0$) we may rewrite Eq. (\ref{Sk2})  as
 \begin{equation}
  S_\ell(k)=\prod_{m=1}^{N_\sigma}\frac{k+\sigma_m}{k-\sigma_m}
	    \prod_{n=1}^{N_\tau}  \frac{k+\tau_n}{k-\tau_n}
            \prod_{i=1}^{N_\xi }  \frac{k+\xi_i}{k-\xi_i}
\label{Sk3}
\end{equation}
or adopting the same symbols for all poles  and roots
\begin{equation}
	 S_\ell(k)=\prod_{m=1}^M\frac{k+a_m}{k-a_m}\,, \qquad M=2N
\label{Sk4}
\end{equation}
It is clear that the number of poles $N_\xi$, $N_\sigma$,
and $N_\tau$ for the $\xi$, $\sigma$, and $\tau$ poles are related
by  $N_\xi=N_\sigma+N_\tau$.

 The form (\ref{Sk4}) is the one used to parametrized the $S$--function 
or equivalently the phase-shifts $\delta$.
This can be easily achieved by rewriting (\ref{Sk4})  in the form
\begin{equation}
	 S_\ell(k)=\frac{1+\sum_{m=1}^MA_mk^m}{1+\sum_{m=1}^M(-1)^mA_mk^m}
\label{Sk5}
\end{equation}
or
\begin{equation}
	 S_\ell(k)+S_\ell(k)\sum_{m=1}^M (-1)^m A_mk^m = 1
		+\sum_{m=1}^MA_mk^m
\label{Sk6}
\end{equation}
from which the $A_m$ are evaluated by choosing  $M$ different
values $S_\ell(k_i)$. The poles $a_i$ are then determined as the zeros
of the polynomial in the denominator of Eq. (\ref{Sk5}).

Once the $a_i$ are found, one has to extract the $\alpha$'s
and $\beta$'s and their distribution on the $k$--plane. In the
absence of bound states we have
$$
      	N_\xi=N_\sigma
$$$$
      	M\ =N_\sigma+N_\xi
$$$$
   	N_\tau=0
$$
from which
\begin{equation}
   	N_\xi=N_\sigma=\frac{M}{2}
\label{nobound}
\end{equation}

In practice one  varies $M$ and the position $k_i$ until
a good fitting of $S_\ell(k)$ at all energies is achieved that fulfills
the condition (\ref{nobound}). In the presence of bound states
the sorting is quite tricky. For example, in the presence of
one bound state one has
$$
      	N_\xi=N_\sigma
$$$$
      	M\ =N_\sigma+N_\xi
$$$$
   	N_\tau=1
$$
Thus
\begin{equation}
   	N_\xi=\frac{M}{2}\,, \qquad N_\sigma=\frac{M-2}{2}\,,
                          \qquad N_\tau=1\,.
\label{withbound}
\end{equation}
The number of poles in the upper $k$--plane $N_u$ is given by
$$
	N_u=N_\xi+N_\tau= \frac{M}{2}+1
$$
while  the number of poles  in the lower $k$--plane $N_l$ by
$$
	N_l=N_\sigma= \frac{M}{2}-1 \,.
$$
Therefore
$$
	N_u=N_l+2\,.
$$
The same argumentation can be used for any number of bound states $N_\tau$
to obtain
$$
	N_u=N_l+2N_\tau\,.
$$
This condition must be satisfied for a correct construction of a potential.

\newpage


\begin{table}[t]
\caption{Comparison of the parameters of the first resonances generated
in each partial wave by the exact potential (19) and by
the approximate analytical form (20).}
\begin{center}
\begin{tabular}{|c|c|c|c|l|}
\hline
&\multicolumn{2}{c|}{exact potential}&
 \multicolumn{2}{c|}{fitted potential}\\
\hline
 $\ \ell\ $ &  $E_r\ ({\rm MeV})\phantom{--} $ &
 $\Gamma\ ({\rm MeV})^{\mathstrut}_{\mathstrut}\phantom{--} $ &
 $E_r\ ({\rm MeV})\phantom{--} $ &
 $\Gamma\ ({\rm MeV})^{\mathstrut}_{\mathstrut}\phantom{--} $\\
\hline
	0 & 2.252380731 & 0.000118256 & 2.2520 & 0.0001179\\
\hline
	1 & 0.807634844 & 0.000000110 & 0.8082 & 0.000000110\\
\hline
	2 & 2.384151637 & 0.000082862 & 2.3843 & 0.000082812\\
\hline
	3 & 1.009031953 & 0.000000046 & 1.0095 & 0.000000046\\
\hline
	4 & 2.676524768 & 0.000027824 & 2.6768 & 0.00002783\\
\hline
\end{tabular}
\end{center}
\label{recov}
\end{table}
\begin{table}[b]
\caption{Spectral points generated by the $\alpha-\alpha$ potential
proposed by Buck {\em et al.} [23].}
\begin{center}
\begin{tabular}{|c|c|c|c|c|}
\hline
&\multicolumn{2}{c|}{momentum (fm$^{-1}$)}
&\multicolumn{2}{c|}{energy (MeV)}\\
\hline
$\ \ell\ $ & $\ {\rm Re\,}(k)\ $ & $\ {\rm Im\,}(k)\ $ &
$\ {\rm Re\,}(E)\ $ & $\ \Gamma\ $ \\
\hline
	0 & $0$           & $2.63671$         & $\ -72.62600\ $ & 0 \\
	0 & $0$           & $1.56602$         & $-25.61905$ & 0 \\
	2 & $0$           & $1.45123$         & $-22.00095$ & 0 \\
	0 & $0.13098$     & $\ -0.00017340\ $ & $0.17922$   & $\ 0.00094905\ $ \\
	2 & $\ 0.53895\ $ & $-0.061203$       & $2.99519$   & $1.37832$    \\
	4 & $1.07513$     & $-0.084009$       & $12.00121$  & $3.77407$    \\
	6 & $1.85894$     & $-0.48994$        & $33.59171$  & $38.05717$   \\
\hline
\end{tabular}
\end{center}
\label{bucktable}
\end{table}
\newpage

\newpage

\newpage


\begin{thebibliography}{99}
\bibitem{Hod94}
	P.E. Hodgson,
       {\em The Nucleon Optical Potential}, (World Scientific, 1994).
\bibitem{RGM}
      	K. Wildermuth and Y. C. Tang, {\em A unified theory of the
         Nucleus}, in {\it Clustering Phenomena of the Nucleus}
	Vol. 1, (Brounsceving, Viliveg, 1977).
\bibitem{F-ELP}
	H. Fiedeldey,   Nucl. Phys. {\bf 77}, 149 (1966).
\bibitem{FS-ELP}
	H. Fiedeldey and S. A. Sofianos,
        Z. Phys,  {\bf A311}, 339 (1983).
\bibitem{S-ELP}
         S. A. Sofianos,  Phys. Rev. C{\bf 35}, 894 (1987).
\bibitem{Marchenko}
    	Z. S.  Agranovich and V. A.  Marchenko,
    	{\it The Inverse Problem of Scattering Theory},
    	(Gordon \& Breach, New York), 1964
\bibitem{Chadan}
	K. Chadan and P. C. Sabatier
	{\em Inverse Problems in Quantum Scattering Theory},
        (Springer, New York), 1989.
\bibitem{Newton}
     	R. G. Newton, {\em Scattering Theory of Waves and Particle},
   	2nd ed. (Springer, New York, 1982).
\bibitem{res1}
	 S. A. Rakityansky, S. A. Sofianos, and K. Amos,
          Il Nuovo Cim. B {\bf 111}, 363 (1996).
\bibitem{res2}
	S. A. Sofianos and  S. A. Rakityansky, J. Phys. A: Math. Gen. 
        {\bf 30}, 3725 (1997).
\bibitem{res3}
	 S. A. Rakityansky and S. A. Sofianos, J. Phys. A: Math. Gen,
         {\bf 31},  5149 (1998).
\bibitem{res4}
	 S. A. Sofianos, S. A. Rakityansky,  and S. E. Massen,
	 to appear in Phys. Rev. A (1999), available from the
	 Los Alamos e-print archive as {\bf nucl-th/9901023}.
\bibitem{Morse} 
        P. M. Morse and H. Feshbach, {\em Methods of theoretical
	physics}, (McGrow-Hill Book Company, New York, 1953) p. 390.
\bibitem{Kuku} 
	V. I. Kukulin, V. N. Pomerantsev, A. Faessler, A. J. Buchmann,
	E. M. Tursunov, Phys. Rev. C {\bf 57}, 535 (1998).
\bibitem{Bay87}
	D. Baye, Phys. Rev. Lett. {\bf 58}, 2738 (1987); J. Phys. A
	{\bf 20}, 5529 (1987).
\bibitem{Sof90}
     	S. A. Sofianos, A. Papastylianos, H. Fiedeldy, and
         E. O. Alt, Phys. Rev. C {\bf 42}, R506, (1990).
\bibitem{Suk85}
	C. V. Sukumar, J. Phys. A {\bf 18}, L57 (1985);
         J. Phys. A {\bf 18}, 2937 (1985);
        J. Phys. A {\bf 18}, {\bf 18}, 2937 (1985).
\bibitem{Fied90}
     	H. Fiedeldy, S. A. Sofianos, A. Papastylianos
        K. Amos, and L. J. Allen,  Phys. Rev. C {\bf 42}, 411, (1990).
\bibitem{MERLIN}
        D. G. Papageorgiou, I. N. Demetropoulos, and I. E. Lagaris,
        Comp. Phys. Comm., {\bf 109}, 227 (1998).
\bibitem{Dubo}
        S. B. Dubovichenko and A. V. Dzhazairov-Kakhramanov,
  	Sov. J. Nucl. Phys. {\bf 51}971 (1990).
\bibitem{Buck77}
     	B. Buck, H. Friedrich, and C. Wheatley, Nucl. Phys.  {\bf A275},
     	246 (1977).
\end{thebibliography}
\end{document}